\documentclass[twocolumn,prl,aps,showpacs]{revtex4}

\usepackage{subfigure}
\usepackage{psfrag,graphicx}
\usepackage{dcolumn}
\usepackage{amsmath,amssymb}
\usepackage{bm}
\usepackage[dvips]{color}
\usepackage{overpic}
\definecolor{mygrey}{gray}{0.35}
\definecolor{mygreen}{rgb}{0.85,1,0.9}
\definecolor{myzard}{cmyk}{0,0,0.05,0}
\definecolor{mywhite}{rgb}{1,1,1}
\definecolor{myred}{rgb}{1,0,0}

 \def\ee{\mathord{\rm e}}
 
 \def\ii{\mathord{\rm i}}

\def\half{\textstyle\frac{1}{2}}

\def\fourth{\textstyle\frac{1}{4}}

\renewcommand{\ii}{{\rm i}}
\renewcommand{\ee}{{\rm e}}
 \newcommand{\ket}[1]{|#1\rangle}
 \newcommand{\bra}[1]{\langle #1|}

\begin{document}

\title[Short Title]{ Competing many-body interactions in systems of trapped ions }

\author{A. Bermudez$^{1}$, D. Porras$^{1,2}$, and M. A. Martin-Delgado$^{1}$}

\affiliation{ $^1$Departamento de F\'{i}sica Te\'orica I,
Universidad Complutense, 28040 Madrid, Spain \\$^2$Max-Planck-Institut f\"{u}r Quantenoptik, Hans-Kopfermann-Strasse 1, 85748 Garching, Germany }

\begin{abstract}
We propose and theoretically analyse an experimental configuration
in which lasers induce 3-spin interactions between trapped
ions.
By properly choosing the intensities and frequencies of the lasers, 3-spin couplings
may be dominant or comparable to 2-spin terms and magnetic
fields.
In this way, trapped ions can be used to
study exotic quantum phases which do not have a counterpart in
nature.
We study the conditions for the validity of the effective
3-spin Hamiltonian, and predict qualitatively the quantum phase
diagram of the system.
\end{abstract}

\pacs{03.67.Ac, 42.50.Wk, 75.10.Jm, 03.67.-a}

\maketitle

Most theoretical models in condensed matter physics rely on two-body
interactions to describe a wide variety of phenomena, like for example,
quantum phase transitions \cite{sachdev_book}.
However, the presence of many-body interactions leads to exotic
quantum effects, such as the
existence of topologically ordered phases with anyonic excitations,
which cannot be generally induced by pairwise couplings \cite{wen_book}.
The beauty and complexity of those exotic models have
motivated a good deal of recent theoretical work.
Unfortunately, theory is still ahead of experimental
implementations in this field, since  many-body interactions are usually
negligible in most  systems found in nature.

In this Letter, we show that 3-body effective spin interactions may be implemented with trapped ions, thus opening a new avenue of research beyond the
traditional paradigm of condensed matter physics.
The application of systems of trapped ions to the quantum simulation
\cite{feynman} of spin models has been theoretically analyzed,
with a focus on conventional models with 2-body interactions \cite{porras_cirac}.
Furthermore, a recent proof-of-principle
experiment has confirmed the validity of this idea \cite{porras_schatz}.
By studying the implementation of
3-spin couplings, we show here that trapped ions can also be used to
engineer new quantum states of matter, with a phenomenology that
does not have a counterpart in nature.
Recently, 3-body interactions have been theoretically analyzed in
systems of ultracold atoms in
triangular lattices \cite{plenio_pachos} and cold polar molecules \cite{zoller}.
Our proposal would benefit from the advantages of experiments
with trapped ions, such as measurement and manipulation at the single
particle level.

To sketch our idea, let us consider a system of spins and a spin operator 
$\sigma^z = \sum_j \sigma^z_j$.
The spins are coupled to a bosonic mode ($H_0 = \delta a^\dagger a$) 
by both a linear, $H_d = F \sigma^z (a + a^\dagger)$, and a quadratic squeezing
term, $H_s = M \sigma^z (a^2 + {a^\dagger}^2)$.
After the adiabatic elimination of the mode, the lowest order
energy corrections include 2-body ($\sim (F^2/\delta) \sigma_j^z\sigma_k^z, (M^2/\delta) \sigma_j^z\sigma_k^z$), 
and 3-body ($\sim (M F^2/\delta^2)\sigma_j^z\sigma_k^z\sigma_l^z$) interactions.
In our scheme, this idea is realized in a system of ions in a linear
array of microtraps \cite{chiaverini,chuang}.
The internal states of the ions play the role of the effective spins.
Two off-resonant laser beams are tuned such that they induce linear
and squeezing couplings between internal states and the collective
vibrational modes of the chain. 
A third off-resonant beam induces 
an auxiliary linear coupling to partially cancel 2-body interactions, 
and thus, to tune the relative strength of the 2- and 3-spin
couplings. In this way, experiments may reveal quantum effects beyond the usual
pairwise-induced correlations, such as a quantum tricritical point,
and a quantum phase with a 3-body order parameter.
\begin{figure}[!hbp]
\centering
\begin{overpic}[width=7.0cm]{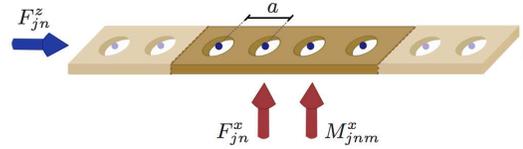}
\end{overpic}
\caption{ Linear  array of microtraps subjected to three laser beams along the axial and radial directions.\vspace{-1ex}}
\label{scheme_microtraps}
\end{figure}

\begin{figure}[!hbp]
\centering
\subfigure[ \hspace{1ex} Radial modes]{
\begin{overpic}[width=4.1cm]{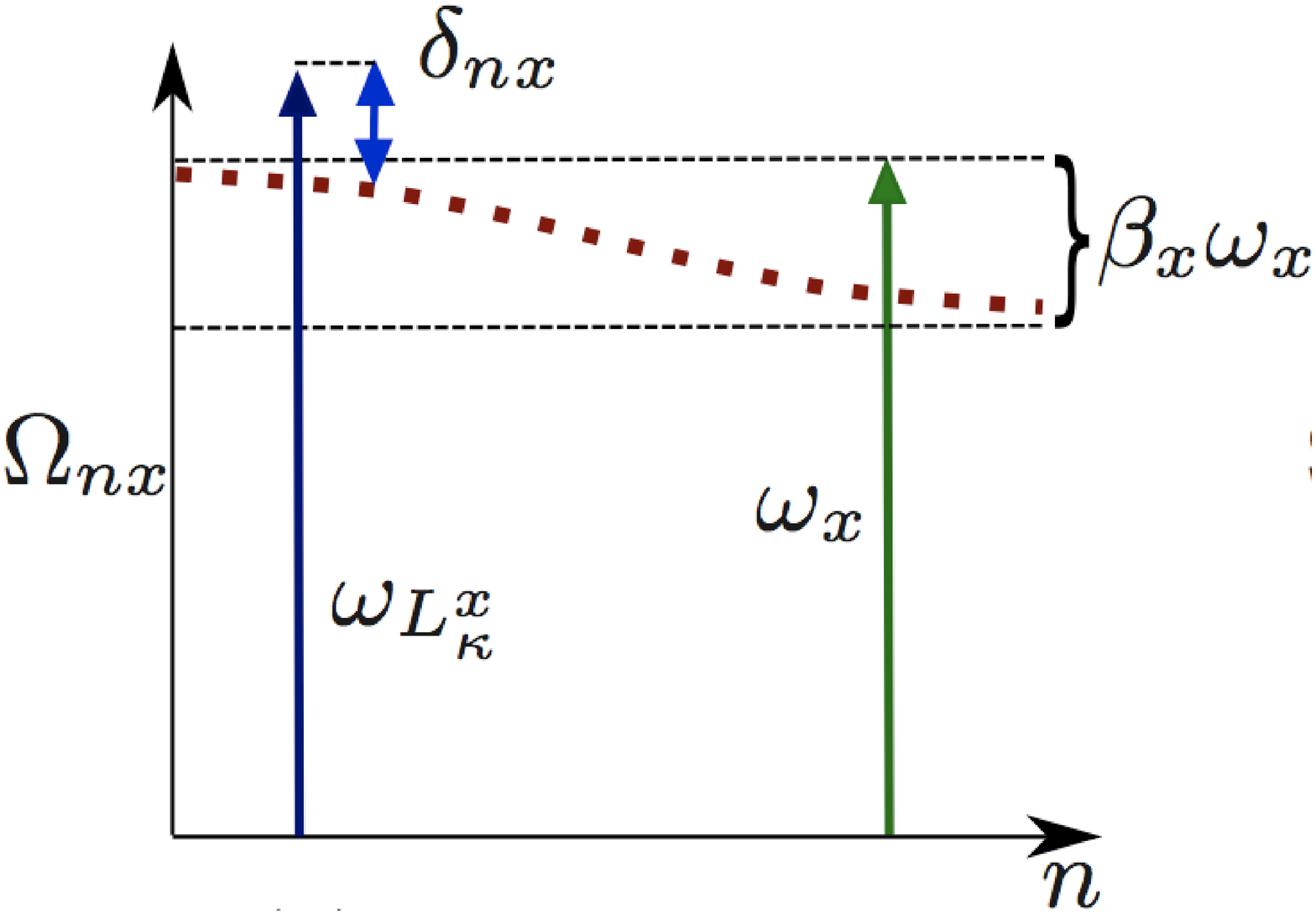}
\label{radial_mode}

\end{overpic}
}\hspace{-2ex}
\subfigure[ \hspace{1ex} Axial modes]{
\begin{overpic}[width=4.1cm]{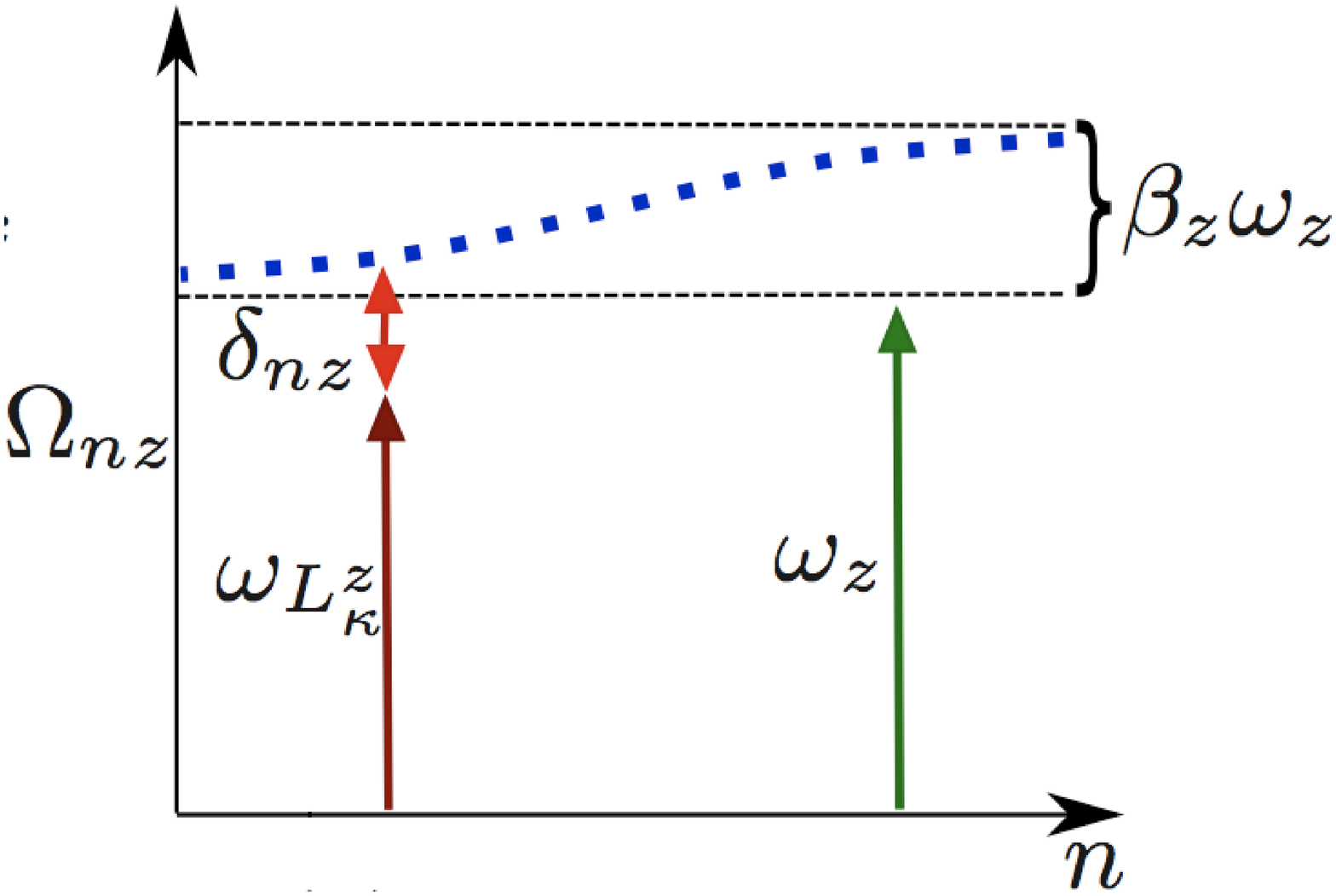}
\label{axial_mode}
\end{overpic}
}
\caption{Phonon  frequencies $\Omega_{n\alpha}$ of (a) radial and (b) axial modes in the stiff limit $\beta_{\alpha}\ll
  1$, where the corrections due to the Coulomb energy $\half c_{\alpha}\beta_{\alpha}\omega_{\alpha}\mathcal{V}_n$ have opposite signs for transverse $c_{x,y}=1$ or longitudinal $c_z=-2$ phonons. The coupling of spin-dependent  dipole forces to
   vibrational modes is also shown: (a) In the radial stiff limit, we can simultaneously blue-detune every mode to the
  first-sideband $\omega_{L_{1}^x}\gtrsim\Omega_{nx}$, 
or to the  second-sideband $\omega_{L_{2}^x}\gtrsim2\Omega_{nx}$. 
(b) In the axial stiff limit, every mode shall be red-detuned to the first-sideband $\omega_{L_{3}^z}\lesssim\Omega_{nz}$.\vspace{-2ex}}
\end{figure}

We start by considering $N$ ions of mass $m$ and
charge $e$, confined along a one-dimensional array of microtraps,
with lattice spacing $a$ and trapping frequencies $\omega_{\alpha}$
with $\alpha=x, y, z$ (see fig.~\ref{scheme_microtraps}).
We assume that the ions have two internal hyperfine ground state
levels ($ \ket{ \! \! \uparrow }$, $|\!\!\downarrow \rangle$), and their motion
can be accurately described in terms of collective
quantized vibrations (i.e. phonons). Therefore, the system Hamiltonian ($\hbar = 1$) becomes
\begin{equation}
\label{free_hamiltonian}
H_0=-\sum_{j=1}^N h\sigma^x_j +
\sum_{n=1}^N\sum_{\alpha=x,y,z}
\Omega_{n\alpha}a^{\dagger}_{n\alpha}a_{n\alpha}.
\end{equation}
Here, $h$ is an effective magnetic field
induced by a laser or microwave field coupled to the
internal transition, $\sigma^\alpha_j$ are the Pauli
matrices corresponding to each ion, $a_{n\alpha}^{\dagger}$ ($a_{n\alpha}$) are the phonon creation
(annihilation) operators, and $\Omega_{n\alpha}=\omega_{\alpha}(1+c_{\alpha}\beta_{\alpha}\mathcal{V}_n)^{\tiny{\frac{1}{2}}}$ denotes the
$n$-th normal mode frequency along the $\alpha$-axis. These frequencies are obtained by diagonalizing  $\mathcal{V}_n=\sum_{jk}{\cal M}_{j n}V_{jk}{\cal M}_{k n}$,  where $V_{jk}=\frac{1}{|j-k|^3}(1-\delta_{jk})-\sum_{l\neq j}\frac{1}{|j-l|^3}\delta_{jk}$ is the Coulomb interaction in the limit of  small vibrations, ${\cal M}_{j n}$ are the normal mode wavefunctions, and $\beta_{\alpha}=e^2/4\pi\epsilon_0m\omega_{\alpha}^2a^3$, $c_{x,y}=1$, $c_z=-2$. Note that in the ''stiff'' limit $\beta_{\alpha}\ll1$, the trapping potential is much larger than the
Coulomb repulsion, and thus  vibrational
frequencies are restricted to a narrow band of width $\half c_{\alpha}\beta_{\alpha}\omega_{\alpha}$ around the trapping
frequencies $\omega_\alpha$ (see figs.~\ref{radial_mode} and \ref{axial_mode}).

The ion chain will be subjected to
three spin-dependent dipole forces $H_I = H_{L^x_1}+H_{L^{x}_{2}}+H_{L^{z}_{3}}$, such that the Hamiltonian becomes
$H = H_0 + H_I$. Here, the most general definition of a dipole force along direction $\alpha$ is 
\begin{equation}
\label{walking_wave}
H_{L_{\kappa}^{\alpha}}=
\frac{\Omega_{L_{\kappa}^{\alpha}}}{2}\sum_{j=1}^N
\left(\ee^{i(k_{L_{\kappa}^{\alpha}}{r}_{j\alpha}-\omega_{L_{\kappa}^{\alpha}} t)}+\text{h.c.}\right)\sigma_j^z,
\end{equation}
which is induced by a pair of lasers in a Raman configuration,
such that $\omega_{L_{\kappa}^{\alpha}}$ is the detuning between the Raman beams,
$\textbf{k}_{L_{\kappa}^{\alpha}}$ is the difference between the laser
wavevectors~\cite{leibfried}, and $\Omega_{L_{\kappa}^{\alpha}}$ is the
two-photon Rabi frequency ~\cite{note_standing_wave}. The position of the ions in~\eqref{walking_wave} 
can be expressed in terms of normal modes, 
$r_{j \alpha} = r^0_{j \alpha} 
+ \sum_{n} 
\textstyle{\frac{{\cal M}_{j n}}{\sqrt{2 m \Omega_{n \alpha}}}}(a^\dagger_{n \alpha} + a_{n \alpha})$, where 
$\textbf{r}^0_j = x_0\textbf{e}_x+ja\textbf{e}_z$ are the ion equilibrium
positions. Let us stress that in the stiff regime, $\beta_{\alpha}\ll1$, it is possible to tune 
 the dipole forces to a certain sideband of every normal
mode of the ion chain, and thus couple each spin to the whole ensemble of vibrational phonons.

The main effect of 
$H_{L^x_1}$ and $H_{L^x_2}$ is to induce conditional linear and squeezing
terms, which will produce 2- and 3-spin couplings, respectively.
This is achieved by tuning the dipole forces
to the first and second blue sidebands ($\omega_{L^x_1} \gtrsim \Omega_{nx}$,
$\omega_{L^x_2} = 2 \omega_{L^x_1} \gtrsim 2 \Omega_{nx}$, see 
fig.~\ref{radial_mode}). Besides, the role of $H_{L^z_3}$ is  to generate
2-body couplings with an opposite sign as those induced by $H_{L^x_1}$, and thus to partially cancel 2-spin interactions in favour of 3-body effects. 
As discussed below, the optimal  screening is achieved by red-detuning
this force to the first axial-sideband ($\omega_{L_3^z}
\lesssim \Omega_{n z}$, see fig.~\ref{axial_mode})).
In the limit of resolved sidebands
$|\delta^{\alpha}_{n\kappa}|=|\Omega_{n\alpha}-\omega_{L^{\alpha}_{\kappa}}|\ll\Omega_{n\alpha}$,
and weak couplings  $\Omega_{L^{\alpha}_{\kappa}}\ll\Omega_{n\alpha}$, we
get a time-dependent interaction Hamiltonian, $H_I(t) = H_d(t) + H_s(t)$, with
\begin{equation}
\label{many_modes_walking_wave}
\begin{split}
H_{d}(t)&=\sum_{j,n}F^{x}_{jn}(t)a_{nx}^{\dagger}\sigma_j^z +\sum_{j,n}F^{z}_{jn}(t)a_{nz}^{\dagger}\sigma_j^z
+\text{h.c.}\\
H_{s}(t)&=\sum_{j,n,m}M^{x}_{jnm}(t)a_{nx}^{\dagger}a_{mx}^{\dagger}\sigma_j^z+\text{h.c.},
\end{split}
\end{equation}
where
we have introduced the coupling strengths
$F^{x}_{jn}(t)=
\ii(\Omega_{L^x_1}\eta_{n1}^{x}\sin\omega_{L^x_1}t
+\Omega_{L^x_2}\eta_{n2}^{x}\sin\omega_{L^x_2}t)
\mathcal{M}_{jn}\ee^{\ii\phi_{x}}$, 
and
$F^{z}_{jn}(t)=\ii(\Omega_{L^z_3}\eta_{n3}^{z}\sin\omega_{L^z_3}t)\mathcal{M}_{jn}\ee^{\ii \phi_{jz}}$ for the contributions arising from the first-sideband, and analogously
$M^{x}_{jnm}(t)=\frac{1}{2}(\Omega_{L^x_2}\cos\omega_{L^x_2}t)\eta_{n2}^x\eta_{m2}^x\mathcal{M}_{jn}\mathcal{M}_{jm}\ee^{\ii\phi_{x}}$ for the  second-sideband.
Let us note that these couplings are switched on adiabatically, and  depend upon the laser
intensities through
$\Omega_{L^{\alpha}_{\kappa}}$,
the Lamb-Dicke parameters
$\eta^{\alpha}_{n\kappa}=k_{L^{\alpha}_{\kappa}}/\sqrt{2m\Omega_{n\alpha}}$,
and the relative phases
$\phi_{j\kappa}^{\alpha}=k_{L^{\alpha}_{\kappa}}r^0_{j\alpha}$, assuming
for simplicity  $\phi^x_{j 1} = \phi^x_{j 2} = \phi^x$.

To get the explicit form of the 2- and 3-spin couplings, we first
note that $H_0 + H_d(t)$ describes a set of forced harmonic
oscillators. If we consider that dipole forces are adiabatically
switched on, 
and $t \gg 1/(\Omega_{n \alpha} -
\Omega_{L^\kappa_\alpha})$, the evolution operator corresponding to $H_0 + H_d(t)$ is 
$U_d(t) = \ee^{-\ii t H_0}\ee^{-\ii t H^{(2)}_{\text{eff}}}\ee^{S}$, with
\begin{equation}
\label{2_spin}
\begin{split}
&H_{\text{eff}}^{(2)}=\sum_{j>k}J_{jk}^{(2)}\sigma_j^z\sigma_k^z, \hspace{2ex}J_{jk}^{(2)}=\left(J^{x}_{jk}+J^{z}_{jk}\right)\\
&J_{jk}^{\alpha}=
\sum_{n\kappa\lambda}\left({\frac{\Omega_{L^{\alpha}_{\kappa}}}{2}}\right)^2{\frac{(\eta^{\alpha}_{n\kappa})^2\mathcal{M}_{jn}\mathcal{M}_{kn}}{\Omega_{n\alpha}-\lambda\omega_{L^{\alpha}_{\kappa}}}},
\end{split}
\end{equation}
where $\lambda=\pm1$ also considers non-resonant terms, and 
\begin{equation}
\label{canonical_transf}
S = \sum_{nj\alpha}\xi_{jn}^{\alpha} a^{\dagger}_{n\alpha}\sigma_j^z-\text{h.c.},\hspace{0.5ex}\xi_{jn}^{\alpha}=\ii\int_0^{t}d\tau
F^{\alpha}_{jn}(\tau)\ee^{\ii\Omega_{n\alpha}\tau}.
\end{equation}
Let us emphasize here that the effective 2-spin interactions mediated by transverse ($J_{kJ}^{\tiny{x}}>0$), and longitudinal phonons ($J_{jk}^{\tiny{z}}<0$), have opposite signs that interfere and  lead to a partial screening of the two-body interactions. In this way, the effects of a 3-body coupling, which arise as a consequence of the additional non-linearities introduced by 
$H_s(t)$, can be experimentally accessed.
Note that we are extending here the canonical transformation used in
\cite{porras_cirac,wunderlich}, to the time-dependent case, something that may
find a broader application in the context of trapped ion quantum simulation.
Working in the interaction picture with respect to 
$U_d$,  the transformed Hamiltonian becomes
$U_d^{\dagger}  {H}_sU_d = H^{(3)}_{\rm eff} + H_{\rm err}$, where
 \begin{align}
 \label{2_3_spin_ion_microtraps}
&H^{(3)}_{\text{eff}}
=\sum_{j>k>l}J^{(3)}_{jkl}\sigma_j^z\sigma_k^z\sigma_l^z,\\
&J^{(3)}_{jkl}=\sum_{nm\lambda}\textstyle{\left(\frac{\Omega_{L^x_1}}{2}\right)^{\tiny{2}}\frac{\Omega_{L^x_2}\cos{\phi_x}}{2}}\textstyle{\frac{(\eta_{n1}^x\eta_{n2}^x)^2\mathcal{M}_{jn}\mathcal{M}_{kn}\mathcal{M}_{jm}\mathcal{M}_{lm}}{(\Omega_{nx}-\lambda\omega_{L^{x}_{1}})(\Omega_{mx}-\lambda\omega_{L^{x}_{1}})}},\notag
\end{align}
describes the effective 3-spin interactions, and
\begin{equation}
\label{many_modes_residual_spin_phonon}
\begin{split}
H_{\text{err}}=&\sum_{nmjk}M^x_{jnm}(t)\sigma_{j}^z a_{mx}\ee^{-\ii(\Omega_{nx}+\Omega_{mx})t}\big(\delta_{jk}a_{nx}-\\
&\hspace{10ex}-2\xi_{kn}^{x}\sigma_{k}^z\big)+\text{h.c.},
\end{split}
\end{equation}
are the residual non-resonant spin-phonon couplings~\cite{note_error_magnetic_field}. 

The conditions under which
these error terms become negligible, which  are thoroughly described
below, lead to a novel Hamiltonian in quantum magnetism
\begin{equation}
\label{2_3_spin_ion_microtraps}
H_{\text{eff}}=\sum_{j>k}J^{(2)}_{jk}\sigma_j^z\sigma_k^z+\sum_{j>k>l}J^{(3)}_{jkl}\sigma_j^z\sigma_k^z\sigma_l^z-h\sum_j\sigma_j^x,
\end{equation}
where 2- and 3-body interactions contribute. We note here that na\"{i}ve  scalings $J^{(2)}\sim\eta^2$, $J^{(3)}\sim\eta^4$
($\eta \sim 0.2$ in experiments) imply that two-spin couplings shall hinder
3-spin effects. However, as shown in Eq.~\eqref{2_spin}, the contribution to the 2-body terms from $H_{L_x^1}$ ($J_{jk}^{x}>0$) is partially cancelled by that from $H_{L_z^3}$ ($J_{jk}^{z}<0$). This effect becomes optimal if the parameters of the axial dipole force are carefully chosen. In particular, $k_{L^{z}_3}$ should be an integer multiple of  $2\pi/a$, such that the ions sit on positions with the same relative phase of the dipole force. To achieve $J^{(2)}\sim J^{(3)}$,
the laser must be red-detuned, with an intensity $\Omega_{L_z^3}$ and detuning $\delta_z<0$   fulfilling
\begin{equation}
\begin{split}
(\Omega_{L^z_3}^2k_{L^z_3}^2)f_1(\omega_{z},\delta_{z},\beta_{z})&=(\Omega_{L^x_1}^2k_{L^x_1}^2)f_2(\omega_x,\delta_x,\beta_x),\\
(\Omega_{L^z_3}^2k_{L^z_3}^2) f_3(\omega_{z},\delta_{z},\beta_{z})&=(\Omega_{L^x_1}^2k_{L^x_1}^2) f_3(\omega_{x},\delta_{x},\beta_{x})+\\&+(\Omega_{L^x_2}^2k_{L^x_2}^2) f_4(\omega_{x},\delta_{x},\beta_{x}),
\end{split}
\end{equation}
where the functions $f_j(\omega_{\alpha},\delta_{\alpha},\beta_{\alpha})$ are listed in~\cite{constraint_function}.
Furthermore, to obtain short-ranged spin interacting models, the radial stiffness parameter is restricted to $\beta_{\alpha}\sim 0.4\delta_{\alpha}/\omega_{\alpha}$,
where $\delta_{\alpha}=\omega_{\alpha}-\omega_{L_{\alpha}}$ is the detuning with respect to the bare trapping frequency. Under all these experimental constraints, at reach with current technology, we come to expressions for the many-body couplings
\begin{equation}
\label{two_three_spin_couplings}
J^{(2)}_{jk}=J_2\Lambda_{jk},\hspace{0.5ex}
 J^{(3)}_{jkl}= \textstyle{\frac{1}{3}}J_3\left(\Lambda_{jk}\Lambda_{kl}+\Lambda_{kj}\Lambda_{jl}+\Lambda_{jl}\Lambda_{lk}\right),
\end{equation}
where we have introduced the dipolar scaling function
$\Lambda_{jk}=1/|j-k|^3$.
The strength of these couplings is
\begin{equation}
\label{two_three_body_couplings_o_magnitude}
J_2=\frac{\beta_xF_x^2}{|\delta_x|}\left(1+\frac{\chi}{4}\right),\hspace{1.5ex}
J_3=\frac{3\beta_x^2\omega_x^2F_x^2M_x\cos\phi_x}{|\delta_x|^4},
\end{equation}
where $\chi=\beta_z\delta_z f_2(\omega_z,\delta_z,\beta_z)/\beta_x|\delta_x|f_1(\omega_x,\delta_x,\beta_x)$. We have introduced the bare linear and quadratic intensities
by $F_x=\half\Omega_{L_1^x}\eta_x$,
$M_x=\fourth\Omega_{L_2^x}\eta_x^2$,
and  the bare Lamb-Dicke parameter $\eta_x=k_{L_x}/\sqrt{2m\omega_x}$.
Hence, the access to the wavelengths, detunings,
and laser intensities, leads to the controllability
of 2 and 3-body couplings.

 In order to check the viability of this proposal, we should carefully deal with the residual spin-phonon coupling in Eq.~\eqref{many_modes_residual_spin_phonon}, which  contains different non-resonant terms that contribute to the error with $\mathcal{O}\big(M^2_x/\delta_x^2\big)$~\cite{note_cooling} and $\mathcal{O}\big(M^2_xF^2_x/\delta^4_x\big)$. Besides, the canonical transformation in Eq.~\eqref{canonical_transf} leads to an additional error
that scales as $\mathcal{O}\big(F_x^2/\delta_x^2\big)$~\cite{porras_deng}. Hence, a feasible quantum simulation of competing many-body interactions requires the parameters $M_x,F_x$ to be small in comparison to the laser detuning $\delta_x$. At this stage, we can check the viability of this proposal assuming the following available experimental parameters, which fulfill all the constraints above. Considering trapping frequencies $\omega_{x}\sim10\text{MHz}$, laser detunings $\delta_{x}\sim1.25\text{MHz}$, stiffness parameters $\beta_{x}\sim0.05$,  and setting the dipole force intensities  to $F_x\approx M_x\sim0.1\delta_x$, we readily obtain an effective model of competing 2- and 3-spin interactions with $|J_3|\approx J_2\sim 0.6\text{kHz}$
and an error on the order of $E\sim10^{-2}$.  Furthermore, the modification of these parameters within the constraints detailed above,  allows to experimentally access different regimes where $|J_3|/J_2\lessgtr1$ and thus observe the consequences of a competition between the many-body interactions in full glory.
Let us note that one can also go beyond the usual dipolar regime in Eq.~\eqref{two_three_spin_couplings} by considering  $\beta_{x}\omega_{x}/\delta_{x}\geq0.5$. In this case, exotic long range interactions between distant spins arise and offer a exceptional playground where the effects of the range of interactions can be studied. Additionally,  the couplings  can be raised to $|J_3 |\approx J_2\sim1-10\text{kHz}$. In order to achieve this interesting regime, one may relax the trapping frequencies or design the microtrap in such a way that the ion equilibrium distance is lowered.  

The ability to independently tune the couplings
$(J_2, J_3, h)$ offers the opportunity to study novel quantum phases
of interacting spins. To get a qualitative picture,
we consider the Hamiltonian where only the
nearest-neighbour terms of the dipole couplings are kept
\begin{equation}
\label{ising_2_3}
H=
J_2 \sum_j\sigma_j^z\sigma_{j+1}^z
 + J_3\sum_j\sigma_j^z\sigma_{j+1}^z\sigma_{j+2}^z-h\sum_j\sigma_j^x .
\end{equation}
The ground state is determined by the competition between the different terms
in (\ref{ising_2_3}):
$J_2$($>0$) induces anti-ferromagnetic (AF) order,
$J_3$($<0$) will be shown to induce a novel ferrimagnetic (F)
phase, and $h$($>0$) encourages the system to lie in a disordered
paramagnetic (P) regime, where spins are aligned along the $x$-direction.
We start our analysis by considering the following two limits:\\
(i) $J_3 = 0$ ({\it 2-spin quantum Ising model}).
This case is exactly solvable and shows a quantum phase transition at the critical coupling
$J_2^c = h$, between the anti-ferromagnetic doubly degenerate ground state ($J_2 > h$)
$\ket{g_{\text{AF}}}\in\{\ket{\!\!\uparrow\downarrow\cdots\uparrow\downarrow},
\ket{\!\!\downarrow\uparrow\cdots\downarrow\uparrow}\}$,
and the paramagnetic phase
$\ket{g_{\text{P}}}\propto\bigotimes_j(\ket{\!\!\uparrow}_j+\ket{\!\!\downarrow}_j)$
($J_2 < h$) \cite{sachdev_book}.
To quantify the degree of anti-ferromagnetic order, we define the order parameter
$\mathcal{O}_{\text{AF}}(g)
=\frac{-1}{N-1}\sum_{j=1}^{N-1}\bra{g}\sigma_{j}^z\sigma_{j+1}^z\ket{g}$,
which fulfills $\mathcal{O}_{\text{AF}}(g_{\text{P}})=0$,
and $\mathcal{O}_{\text{AF}}(g_{\text{AF}})=1$.\\
(ii) $J_2 = 0$ ({\it 3-spin quantum Ising model}). In this case,
3-spin interactions induce a novel
quantum phase which can be fully characterisecharacterizedd by the order parameter
$\mathcal{O}_{\text{F}}(g)
=\frac{1}{N-2}\sum_{j=1}^{N-2}\bra{g}\sigma_{j}^z\sigma_{j+1}^z\sigma_{j+2}^z\ket{g}$.
The 3-spin Ising model is no longer exactly solvable, but shows
self-duality properties~\cite{Penson_Pfeuty},
something that allows us to locate its critical point at the value $|J_3^c| =
h$. This point separates a phase with a four-fold ferrimagnetic
state,
$\ket{g_{\text{F}}}
\in\{\ket{\!\!\uparrow\uparrow\uparrow\cdots\uparrow\uparrow\uparrow},
\ket{\!\!\uparrow\downarrow\downarrow\cdots\uparrow\downarrow\downarrow},
\ket{\!\!\downarrow\uparrow\downarrow\cdots\downarrow\uparrow\downarrow},
\ket{\!\!\downarrow\downarrow\uparrow\cdots\downarrow\downarrow\uparrow}\}$
($J_3 > h$), from the paramagnetic phase ($|J_3| < h$).
Remarkably, at $J_3^c$ there is a  phase transition which
belongs to f the four-state Potts model  universality class. Hence,
$3$-spin correlations induce an exotic critical behaviour different
from the Ising universality class.

\begin{figure}[!hbp]
\centering
\subfigure[\hspace{1ex}Phase diagram]{
\begin{overpic}[width=3.0cm]{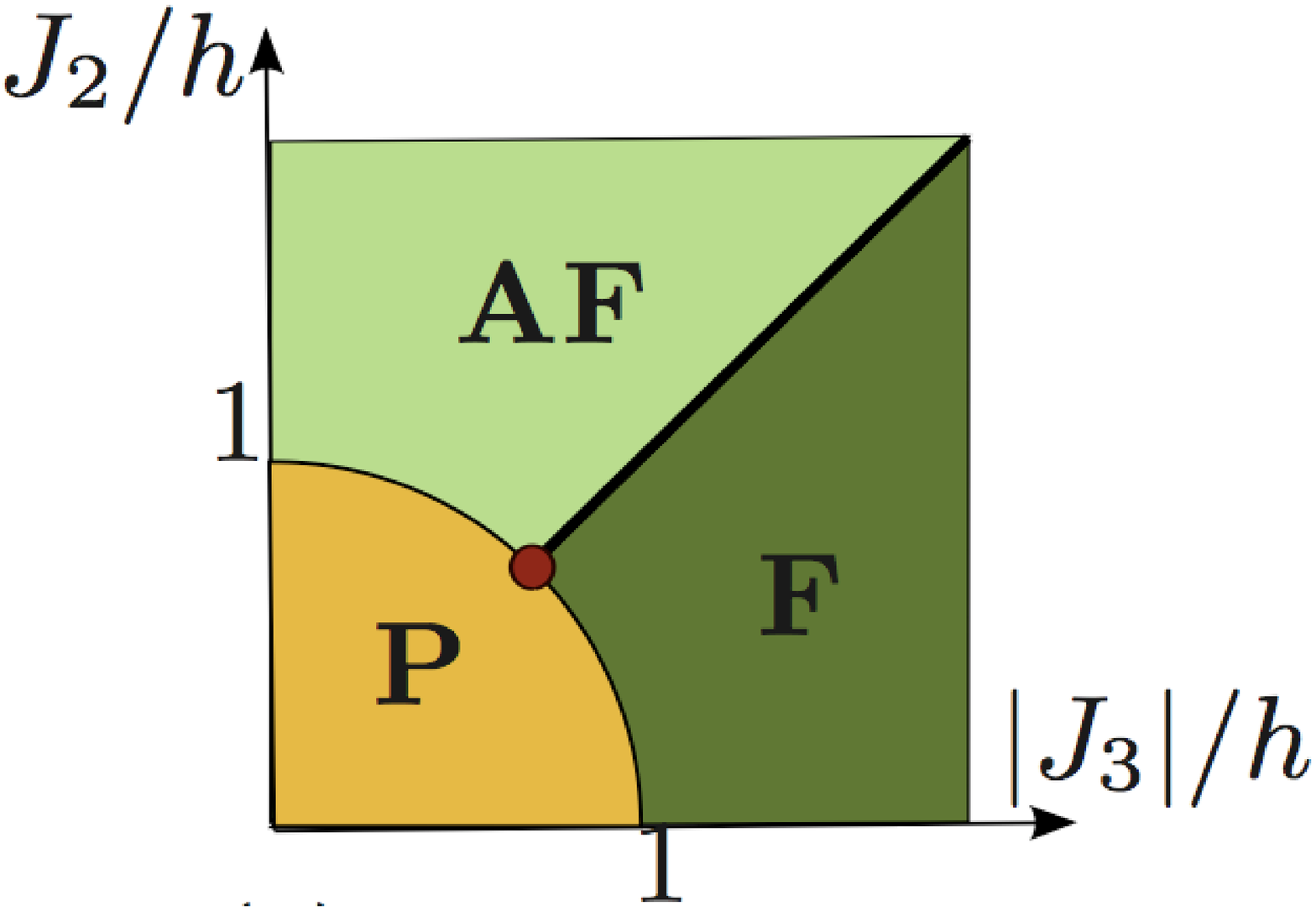} \label{diagrama}

\end{overpic}
}

\subfigure[\hspace{1ex}AF order parameter]{
\begin{overpic}[width=3.0cm]{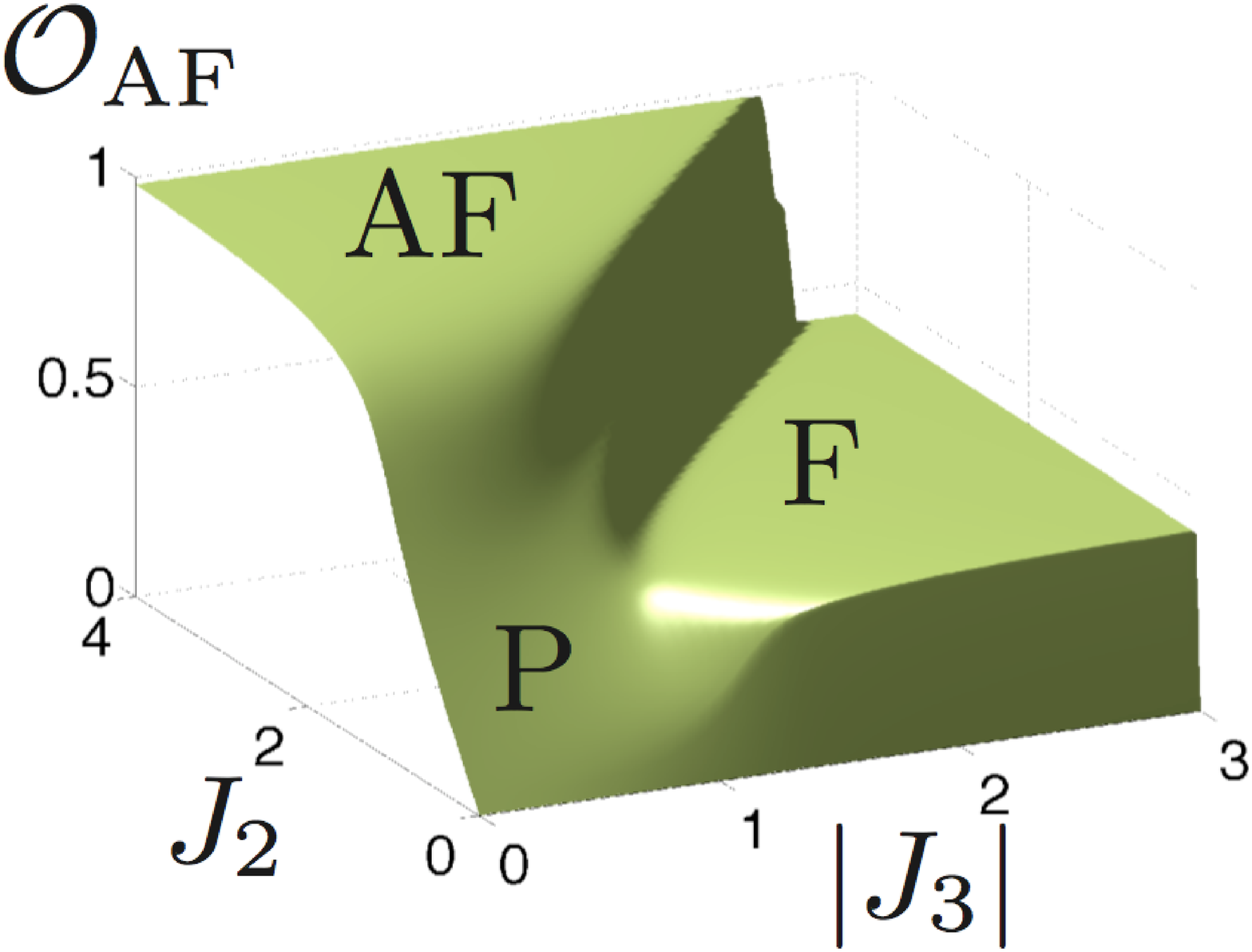} \label{finite_size_zz}

\end{overpic}
\label{AF_op}
}
\subfigure[\hspace{1ex}Ferri parameter]{
\begin{overpic}[width=3.0cm]{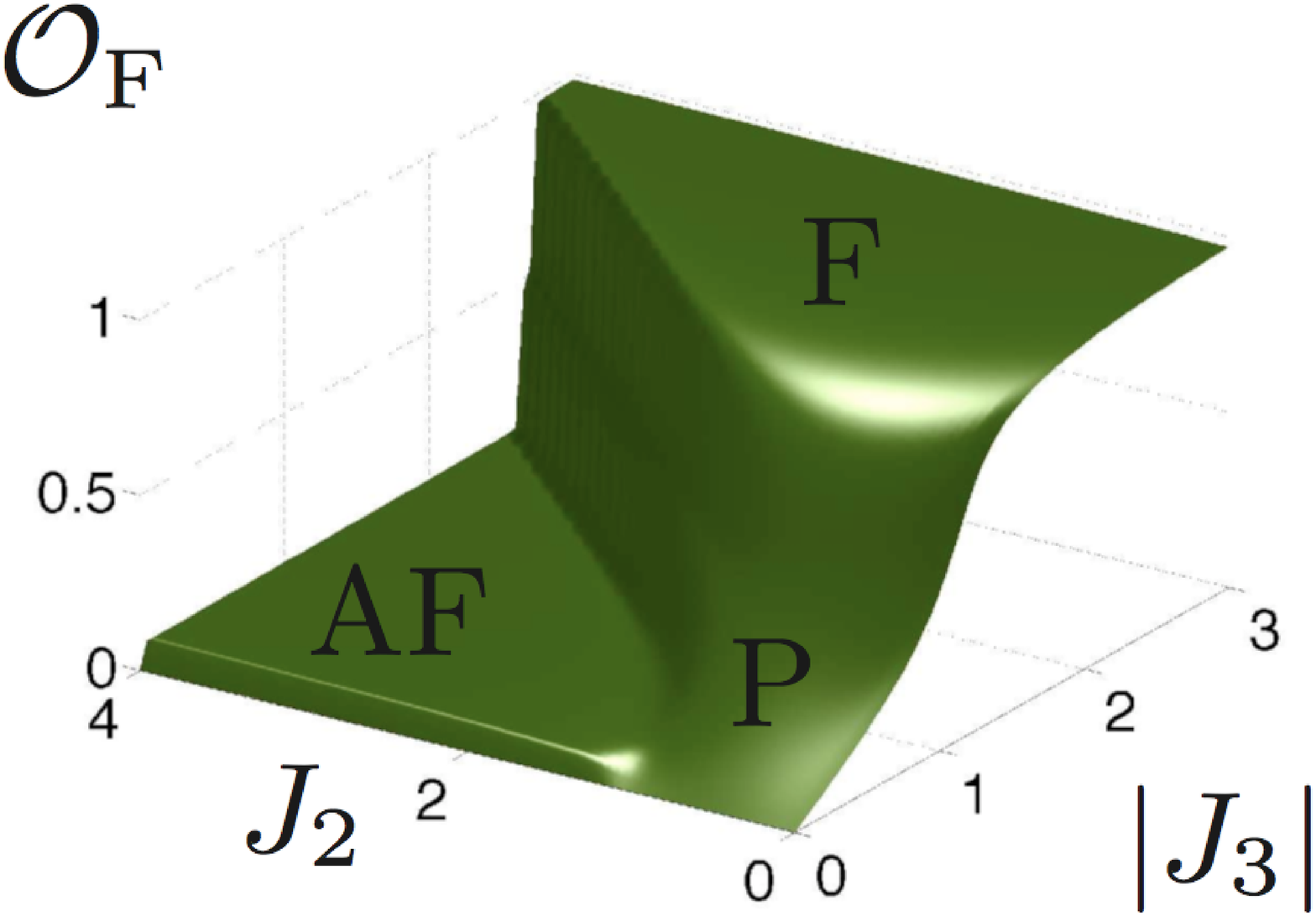} \label{finite_size_zzz}

\end{overpic}
\label{F_op}
}
\caption{(a) Quantum phase diagram with P, AF and F phases that coexist in the tricritical point (in red). Order parameters as a function of the couplings $J_2,J_3$ for a chain with $N=15$ spins: (b) Anti-ferromagnetic order parameter (c) Ferrimagnetic order parameter.\vspace{-2ex}}
\end{figure}

From these limiting regimes, one gets a notion of the
complexity of the model for general 
$(J_2,J_3,h)$. In addition to the critical points studied above, the
system should also hold a quantum phase transition between the
AF and F phases, as well as a tricritical point, at which all the
magnetic orders coexists (see fig.~\ref{diagrama}).
This qualitative picture is supported by the finite-size numerical
calculations presented in figs.~\ref{finite_size_zz} and \ref{finite_size_zzz} . Note that the order
parameters, ${\cal O}_{\rm AF}$, ${\cal O}_{\rm F}$,
may be measured by detecting the photoluminescence from individual ions,
something that amounts to a quantum measurement of $\sigma^z$~\cite{wineland}. It is precisely this ability of performing
highly accurate measurements at the single particle level,
which allows us  to characterize the full quantum phase diagram.

Let us briefly consider the implementation of these spin models using the more conventional linear Paul traps. Although the axial stiff limit cannot be achieved (i.e. the ion chain stability imposes $\beta_z\gtrsim1$), it is still possible to devise effective two- and three-spin interactions. Focusing on the case of three  ions, a similar procedure as that presented for microtraps, but considering a single mode in both radial directions $\alpha=x,y$ would yield 
\begin{equation}
H=J_2(\sigma_1^z\sigma_2^z+\sigma_2^z\sigma_3^z+\sigma_3^z\sigma_1^z)+J_3\sigma_1^z\sigma_2^z\sigma_3^z-h(\sigma_1^x+\sigma_2^x+\sigma_3^x),\notag
\end{equation}
where there is full access to the different couplings $(J_2,J_3,h)$. In this case, we could even switch off the two-body term $J_2=0$ enhancing pure three-spin interactions and vice versa. In the same spirit as~\cite{porras_schatz}, one could perform a proof-of-principle experiment, where an initial separable paramagnetic state $\ket{\text{P}}=\ket{\!\!\rightarrow\rightarrow\rightarrow}$ adiabatically evolves towards an entangled state with different types of ordering. In case we tune $J_2<0,J_3=0$, the evolution would generate GHZ states $\ket{\text{GHZ}}\sim\ket{\!\uparrow\uparrow\uparrow}+\ket{\!\downarrow\downarrow\downarrow}$, whereas for $J_3<0, J_2\gtrsim0$, one generates W-states $\ket{W}\sim\ket{\!\uparrow\downarrow\downarrow}+\ket{\!\downarrow\uparrow\downarrow}+\ket{\!\downarrow\downarrow\uparrow}$, studying thus the two non-equivalent classes of tripartite entanglement. Note  that the Hamiltonians obtained with Paul traps are restricted to mean-field models, and  microtraps should be used  to implement strongly correlated systems.

In conclusion, we have shown that a system of trapped ions can be used to explore the singular phenomenology of spin models with three-body interactions. We have made a realistic experimental proposal, at reach with current technology,
to  access the peculiar phase diagram of an effective Hamiltonian with three-body interactions, characterized by the appearance of  exotic phases without a counterpart in usual condensed matter experiments.

{\it Acknowledgements.} A.B. and M.A.M.D  acknowledge financial support
from the projects FIS2006-04885, CAM-UCM/910758, and INSTANS 2005-2010.  A. B.
acknowledges support from a FPU MEC grant.



\begin{references}
\bibitem{note_standing_wave}
The use of off-resonant walking waves, instead of standing waves~\cite{porras_cirac}, presents the advantage that there is no need to control the position of the ion chain with respect to the applied wave.
\bibitem{note_error_magnetic_field}
There is also an error term proportional to the transverse magnetic
field $H_{\text{err}}=2\ii h\sum_{jn\alpha}\xi_{jn}^{\alpha} a_{n\alpha}^{\dagger}\sigma_j^{y}+\text{h.c.}$ . However, it gives a negligible contribution for the range of parameters that we are interested in $h\approx J_2,J_3$, which leads to insignificant corrections of the order $\mathcal{O}\big(F_x^6/\delta_x^6\big)$.
\bibitem{note_cooling}
In order to neglect the error term quadratic in phonon operators, the ion chain should be cooled down to $\bar{n}<0.1$, which does not require spectral resolution of the different modes (i.e. $k_{\text{B}}T\gg \beta\omega$).
\end{references}
\end{document}